# Coding in the Presence of Semantic Value of Information: Unequal Error Protection Using Poset Decoders*

Marcelo Firer† Luciano Panek‡ Laura Rifo§

**Abstract**

In this work we explore possibilities for coding when information worlds have different (semantic) values. We introduce a *loss function* that expresses the overall performance of a coding scheme for discrete channels and exchange the usual goal of minimizing the error probability to that of minimizing the expected loss. In this environment we explore the possibilities of using poset-decoders to make a message-wise unequal error protection (UEP), where the most valuable information is protected by placing in its proximity information words that differ by small valued information. Similar definitions and results are shortly presented also for signal constellations in Euclidean space.

## 1 Introduction

Since the mid 1990s, some new metrics were introduced in the study of error-correcting codes, mainly metrics determined by a partial order in the set of

*The work of M. Firer was partially supported by FAPESP grant number 2007/56052-8. The work of L. Panek was partially supported by the Fundação Araucária. L. Rifo was partially supported by Proyecto mel-conicyt 81100005.

†M. Firer is with IMECC-UNICAMP, Universidade Estadual de Campinas, CEP 13083-859, Campinas, SP, Brazil (e-mail: mfirer@ime.unicamp.br).

‡L. Panek is with CECE-UNIOESTE, Universidade Estadual do Oeste do Paraná, CEP 85870-900, Foz do Iguaçu, PR, Brazil (e-mail: lucpanek@gmail.com).

§L. L. R. Rifo is with IMECC-UNICAMP, Universidade Estadual de Campinas, CEP 13083-859, Campinas, SP, Brazil (e-mail: lramos@ime.unicamp.br).

positions coordinates of linear codes, called for simplicity just *poset metrics*. The relevance of such metrics is being determined considering channels for which such metric structures are more appropriate than the usual Hamming metric (see [21] and [26]). In this work we generally assume the most usual setting of coding theory, the use of linear codes over *discrete channels* (DC), but introduce a new parameter, the value of the information, that turns poset metrics into a valuable tool for getting decoders with a good performance.

As noted by Claude Shannon at the introduction of his seminal work (see [23]), the "*... semantic aspects of communication are irrelevant to the engineering problem*". In this work, we do not consider the semantic of information, only the possibility of considering its semantic value, something that should be defined by experts in the different fields producing information to be communicated. Considering such a value function, that associates to each information word a non-negative real number, allows us to make a slight but relevant change in one of the main questions that drives coding theory: instead of searching for codes that minimize the quantity of errors, we can look for codes that minimize the overall value of the decoded errors[1].

In this work we establish a general framework for considering value of information in coding theory, presenting first some existence results that open a wide range of new questions. The introduction of expected loss functions generalizes the usual approach of maximum likelihood decoders (ML) and poses a new theoretical goal: instead of looking for a code (with given properties, such as dimension and length) that minimizes the expectation of the number of errors after decoding, we are actually looking for a triple, consisting of a code, the way information is mapped into the code and a decoder that minimizes the expected loss.

To deal with such a larger and difficult goal, we bring into the picture a family of decoders that are in some sense more manageable, decoders that are nearest-neighbor decoders, according to a family of metrics called poset-metrics. Considering those metrics we are able to show, in a general setting, the existence of nearest-neighbor decoders that beats the performance of classical ML decoders. This a posteriori is not surprising since ML decoders answer a different question (minimizing the number of errors). Moreover, considering a particular set of poset, those called hierarchical posets, we are able to move forwards and determine efficient decoding algorithms (see [8] and [20]).

[1]Concerning the question of semantics, we must stress we do not aim to settle a mathematical-theoretical framework that will allow semantical communication, as for example the one being carried by Juba and Sudan ([13]), but we are just assuming that in some sense, a semantical value was attributed to information.

The approach adopted in this work goes somehow in the same direction that has been followed in some recent works. In the decoding process, the use of nearest-neighbor decoders determined by poset-metrics is actually a decoding process that gives unequal error protection for bits (bit-wise UEP), in a similar way as proposed in 1967 by Masnick and Wolf in [16] and since then extensively studied by many authors. Considering unequal error protection of messages (message-wise UEP) instead of bits is the approach adopted by Borade, Nakiboğlu and Zeng in [5], when they consider the necessity of protecting in different ways information that are different in their nature (like data and control messages) or different type of errors (erasures and mis-decoded messages), showing the possibility to achieve the channel capacity exponentially for some more protected bits (by "stealing" the capacity from other bits). The approach given in this work in some sense is more general and combines unequal error protection of messages and bits. Moreover, in the approach adopted in [5], more valuable information is over-protected by assigning larger decoding regions while in this work the approach to message-wise unequal error protection is significantly different: in the former, more valuable information is protected by placing in its neighborhood information with similar semantic value.

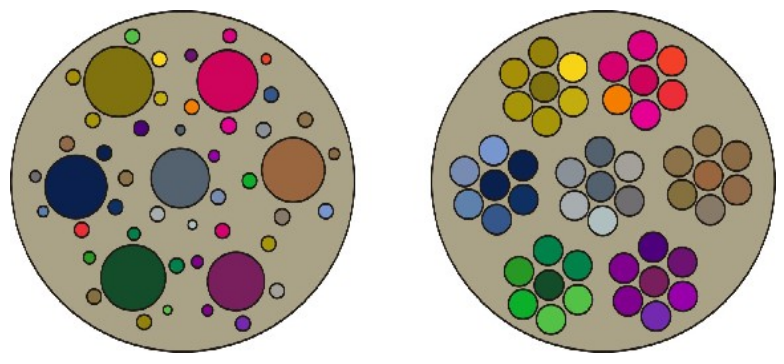

Figure 1: Message-wise UEP and ordered bit-wise UEP.

Also, we do not consider an $[n;k]_q$ linear code $C$ just as a subspace of $\mathbb{F}_q^n$, but as an map $g : I \to \mathbb{F}_q^n$, where $I = \mathbb{F}_q^k$ may be thought as a source code. If we fix such an encoding function $g : I \to \mathbb{F}_q^n$, we are actually distinguishing between $g$ and $g \circ \sigma$, where $\sigma : I \to I$ is any permutation of the information set.

In this sense, we may say are making a joint source-channel coding (JSCC), in the same sense adopted for instance in [9] where some quantized informations are more relevant than others.

As a very simple application, we consider the picture bellow. It is a picture in scale-of-gray encoded in the source with 4 bits of information. The information was encoded as the perfect $[7, 4]_2$ binary Hamming code, one codeword assigned for each pixel.

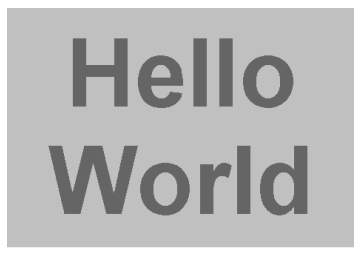

Figure 2: Original picture.

Using a random number generator, an error was created for each of the seven bits of each pixel, with error probability $p = 0.3$. The same received picture was corrected twice, once using usual ML decoder and once using a decoder determined by a given poset (details in Appendix 11), which we call for the moment just a $P$-decoder.

In Figure 3 we can see in a unique different color (purple in the colored version) the pixels that were correctly decoded. The pixels that were incorrectly decoded are presented in the (wrong) color they were decoded. On the left side we see the result for ML decoder and on the right side the result for the $P$-decoder.

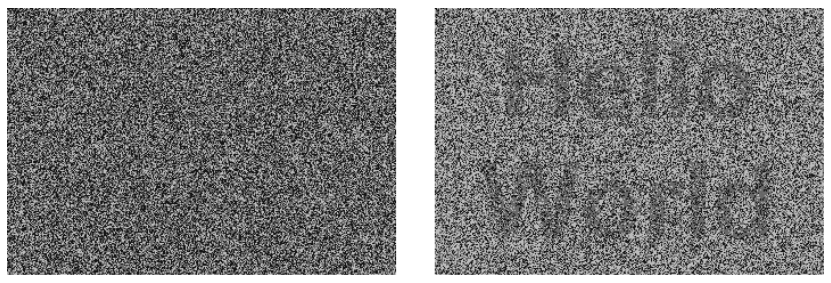

Figure 3: Right corrected pixels are colored with purple; on the left ML decoding and on the right $P$-decoding.

As expected, the picture on the left is much more color homogeneous (purple-like), since using ML to decode with a perfect code minimizes the amount of errors. However, one can identify the pixels to be painted in purple only when having the original picture. When looking at the picture as it was decoded using the two different decoding schemes, one gets a quite different perception:

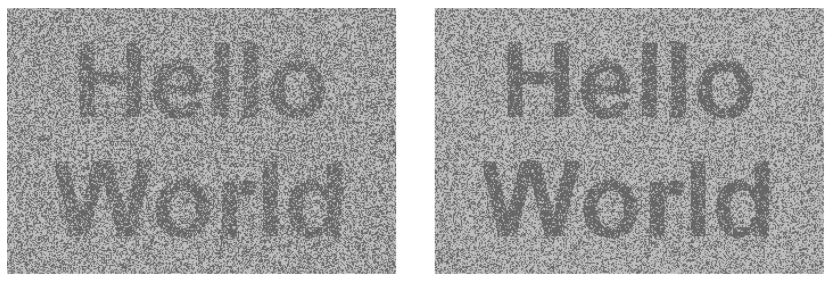

Figure 4: On the left ML decoding and on the right $P$-decoding.

The right-hand image seems to be more sharp, closer to the original picture (Figure 2). This perception about the quality of those decoded pictures is an example of a way of valuing information, in a situation in which each of us, ordinary viewers, may be considered as a kind of expert.

Despite the fact those pictures[2] were made considering a very basic model for encoding a gray-scale palette of colours, they are a good illustration to the main points proposed in this work, including the fact that ML decoding is not always better and poset decoders may give better results.

# 2 Organization

Along this work, we study only linear codes and consider transmission either on a general but not specified discrete channel or sometimes over a discrete symmetric memoryless channel (DSMC). All along the work we assume that every codeword is transmitted with the same probability. Although the fact those restrictions are not essential for most definitions introduced in this work (except the linearity of the codes under consideration) we prefer to restrict ourselves to this context, since actually dealing with more general channels or codes words with different frequencies of transition becomes too intricate for this initial approach.

[2]All the pictures illustrating this section were produced using a software developed by Vanderson Martins do Rosario, a first year undergraduate student at Universidade Estadual de Maringá (UEM) to whom we are in debt.

This work is organized as follows. In Section 3 we recall the basic facts about maximum likelihood (ML), maximum a posteriori (MAP) and nearest-neighbor (NN) decoders. In Section 4 we introduce the main concepts and definitions used in this work: value function, loss function, overall expected loss and Bayes decoder. The main result in this section, Proposition 1, characterizes the expected loss for a DSMC. In Section 5 we describe an analogue of Shannon's theorem for valued information (Theorem 1). After proving those general and structural results, in Section 6 we restrict the set of decoders to the set of NN decoders relative to the poset metrics. Considering the difference between the expected loss of different NN poset decoders we determine a simple condition that assures the existence of two nonempty subsets where one of those NN poset decoders is better than the other (in terms of minimizing the expected loss) and vice-versa (Theorem 2). In Section 7 we present the existence results of this work. The first one states that for any linear code and any ML decoder, there are always value functions for which it is better to use a non-ML decoder (Theorem 4). In another result we show that, for a large infinite family of pairs $(P, Q)$ of posets (called $(I, J)$-decomposable posets) there are codes for which better results (in term of total expected loss) may be attained either by a $P$-NN decoder or a $Q$-NN decoder, according to the value given to each information (Theorem 6). On Section 8, we work with signal constellations in a continuous channel, defining in a similar way what an expected loss function is and showing (Theorem 8) that ML decoders are not necessarily better than other decoders.

This is not a work that gives answers to known questions, but rather a work that aims to show both the convenience and the viability of considering the value of information. Inasmuch, many questions that arise are not answered. Section 9 is devoted to some final remarks and open problems, but there are also some open questions stated along the text, connected to the matter and propositions that made them arose.

In order to make the reading of the work more fluent, we decided to gather most of the proofs in Appendix 10. Finally, in Appendix 11, we present the details about the coding schemes used to produce the pictures presented in the

Introduction.

# 3 Useful Background: ML, MAP and NN Decoders

Let $\mathbb{F}_q^n$ be the linear space of $n$-tuples over a finite field $\mathbb{F}_q$ and $C \subseteq \mathbb{F}_q^n$ be an $[n;k]_q$ *linear code*. Let $d_H(\cdot,\cdot)$ be the usual *Hamming distance*: $d_H(x,y)$ is the number of coordinate positions in which the $x$ and $y$ differ.

A *discrete channel* (DC) over $\mathbb{F}_q$ is characterized by the set of conditional probabilities $\{P(b|a) : a, b \in \mathbb{F}_q\}$ where $P(b|a)$ represents the probability of receiving the symbol $b$ given that the symbol $a$ has been transmitted. We assume that the channel is *memoryless* (DMC), that is

$$P(y|x) = \prod_{i=1}^{n} P(y_i|x_i)$$

where $x = (x_1, \ldots, x_n)$ and $y = (y_1, \ldots, y_n)$ represent $n$ consecutive transmitted and received symbols, respectively. A DMC over $\mathbb{F}_q$ is called *symmetric* (DSMC) with crossover probability $p$ if

$$P(y|x) = (1-p)^{n-d_H(x,y)} \left(\frac{p}{q-1}\right)^{d_H(x,y)}.$$

Considering that a vector $y$ is received through the channel, there are two plausible criteria to decide how to decode it. We can decode $y$ as a codeword $c_y$ such that

$$P(y|c_y) = \max_{c \in C} P(y|c)$$

or we can decode $y$ as a codeword $c_y$ such that

$$P(c_y|y) = \max_{c \in C} P(c|y).$$

The first decoding criterion is called *maximum likelihood decoder* (ML). The second decoding criterion is called *maximum a posteriori decoder* (MAP). Since we are assuming that each codeword $c$ is transmitted with probability $P(c) = \frac{1}{M}$, with $M = q^k$, it follows from Bayes' rule that both ML and MAP decoders coincide with the *nearest-neighbor decoder* (NN) in DSMC:

$$d_H(y, c_y) = \min_{c \in C} d_H(y, c).$$

For each decoding criterion above we may define a (generally not unique) map $a : \mathbb{F}_q^n \to C$ such that

$$d_H(y, a(y)) = \min_{c \in C} d_H(y, c).$$

In general, a *decoding scheme* (or just *decoder*) is just a map

$$a : \mathbb{F}_q^n \to C.$$

It is reasonable to require that $a(c) = c$ for all $c \in C$ and in this situation we call $a$ an *ordinary* or *reasonable decoder.* Let $D(c)$ be the *decision region of* $c$ relative to the decoding scheme $a$:

$$D(c) := a^{-1}(c) = \left\{ y \in \mathbb{F}_q^n : a(y) = c \right\}.$$

The decision regions $D(c)$ of a decoder $a$ determine a partition of $\mathbb{F}_q^n$. Given a decoding scheme, an error occurs if $c$ is sent and the received codeword lies in some decision region $D(c')$, with $c' \neq c$. The *probability of error* is therefore

$$P_e(c) = 1 - \sum_{y \in D(c)} P(y|\, c)$$

where the sum runs over all $y \in D(c)$. As the probability distribution of $C$ is uniform, the *decoding error probability of* $C$ is the average

$$P_e(C) = \frac{1}{M} \sum_{c \in C} P_e(c).$$

We let now $\mathbb{R}_+$ denotes the set of *non-negative real numbers* and consider the map

$$\mu_{\text{0-1}} : C \to \mathbb{R}_+$$

given by

$$\mu_{\text{0-1}}(c) = \begin{cases} 0 & \text{if} \quad c = 0 \\ 1 & \text{if} \quad c \neq 0 \end{cases}.$$

It follows that

$$P_e(C) = \frac{1}{M} \sum_{c \in C} \sum_{y \in \mathbb{F}_q^n} \mu_{\text{0-1}}(a(y) - c)\, P(y|\, c).$$

We remark that at this point it is essential to consider $C$ to be a linear code, in order to ensure that $a(y) - c \in C$. The function $\mu_{\text{0-1}}$ is a characteristic function that only detect decoding errors, but do not distinguish different decoding. We will use the notation $\mu_{0-1}$ for any such function, independently of the code under consideration.

In many real situations, it is reasonable to attribute different values to different codewords, and this is what will be done in this work, considering instead of $\mu_{\text{0-1}}$ value functions that may assume any (non-negative) real value. A typical example of this situation is the transmission of digital images illustrated in the introduction: small variations in the color values of each pixel does not affect the quality perception of the image. This work is inspired by this very common kind of situation.

# 4 Value Functions and Expected Loss

A *value function for a linear code* $C$ is just a map $\mu$ that associates to each codeword a non-negative real number

$$\mu : C \to \mathbb{R}_+,$$

and a *loss function*

$$l : C \times \mathbb{F}_q^n \to \mathbb{R}_+$$

given by $l(c, y) = \mu(a(y) - c)$ gives a measure of the loss when the information $c \in C$ was send and the information $y \in \mathbb{F}_q^n$ was received and decoded as $a(y) \in C$. We remark that, since $C$ is linear, the difference $a(y) - c$ is actually a codeword hence it makes sense to consider the value $\mu(a(y) - c)$. By doing so, we are evaluating the errors that may occur during the process consisting of encoding, transmitting and decoding information. In such a situation it is reasonable to require that $\mu(0) = 0$ (we should not lose anything if everything was right along the process). If this happens, we say the value function is *reasonable*. We make this distinction since we will use some "unreasonable" value functions that proved to be valuable for proving Theorem 4 in section 7.

Given a linear code $C$, a decoder $a$ for $C$, and a value function $\mu : C \to \mathbb{R}_+$, we define the *expected loss* of $a$ relative to $\mu$ and to a received information $y$ to be the average

$$\mathcal{L}_y(a, \mu) = \mathbb{E}(l(c, y)) = \sum_{c \in C} l(c, y) P(c|y) = \sum_{c \in C} \mu(a(y) - c) P(c|y). \quad (1)$$

We define the *overall expected loss* of $a$ as the average of the expected loss for all possible informations $y \in \mathbb{F}_q^n$,

$$\mathbb{E}\left(\mathcal{L}(a,\mu)\right) = \sum_{y\in\mathbb{F}_q^n} \mathcal{L}_y\left(a,\mu\right) P(y),$$

where $P(y) = \sum_c P(c)P(y|c)$ is the probability of receiving $y$. That expression can be rewritten as

$$\mathbb{E}\left(\mathcal{L}(a,\mu)\right) = \sum_{c\in C}\sum_{y\in\mathbb{F}_q^n} \mu\left(a\left(y\right)-c\right) P\left(y|\, c\right) P(c).$$

Making the change of variable $\tau = a\left(y\right) - c$ we have that

$$\mathbb{E}\left(\mathcal{L}(a,\mu)\right) = \sum_{\tau\in C} G_a\left(\tau\right)\mu\left(\tau\right)$$

where

$$G_a\left(\tau\right) = \sum_{y\in\mathbb{F}_q^n} P\left(y|\, a\left(y\right)-\tau\right) P\left(a\left(y\right)-\tau\right). \tag{2}$$

We remark that $\mathbb{E}\left(\mathcal{L}(a,\mu)\right)$ actually depends on $C$ and should be denoted as $\mathbb{E}\left(\mathcal{L}(a,\mu,C)\right)$, but this dependence on $C$ will be omitted when it should not cause any confusion. Also, to shorten the notation and since no confusion may arise we will denote

$$\mathbb{E}\left(a,\mu\right) := \mathbb{E}\left(\mathcal{L}(a,\mu)\right).$$

Moreover, since we are considering the value of information, the total expected loss depends not only on the code itself but also on the way the information is mapped into the code. In other words, we are actually considering a value function $\widetilde{\mu} : I \to \mathbb{R}_+$ where $I$ is a source code. When we say that a code $C$ is given we are assuming that it is given an embedding $g : I \to C \subseteq \mathbb{F}_q^n$ and the function $\mu : C \to \mathbb{R}_+$ is the unique function such that $\mu \circ g = \widetilde{\mu}$.

We say that a decoder $a^*$ is a *Bayes decoder for* $C$ relative to the value function $\mu$ and to the loss function $l$ if for each received information $y \in \mathbb{F}_q^n$ it minimizes the expected loss, i.e.,

$$\mathcal{L}_y\left(a^*,\mu\right) = \min_a \mathcal{L}_y\left(a,\mu\right),$$

where the minimum is taken over the set of all decoders $a$ of $C$.

Given a DSMC with crossover probability $p$ and an $[n;k]_q$ linear code $C$ such that $P(c) = \frac{1}{M}$ for all $c \in C$, we have that

$$P(y|\, c) = (1-p)^{n-d_H(y,c)} \left(\frac{p}{q-1}\right)^{d_H(y,c)}.$$

Thus, in expression (2),

$$G_a(\tau) = z \sum_{y \in \mathbb{F}_q^n} s^{d_H(y, a(y)-\tau)}$$

where

$$z = z(p) := \frac{(1-p)^n}{M}$$

and

$$s = s(p) := \frac{p}{(1-p)(q-1)}.$$

Dropping the multiplicative scaling factor $z$ in $G_a$ we obtain:

**Proposition 1** *For a DSMC we have*

$$\mathbb{E}(a,\mu) = \sum_{\tau \in C} G_a(\tau)\,\mu(\tau)$$

*where*

$$G_a(\tau) = \sum_{y \in \mathbb{F}_q^n} s^{d_H(y, a(y)-\tau)}.$$

# 5 Shannon's Theorem Analogue for $\mathbb{E}(a,\mu)$

Shannon's coding theorem of 1948 (see [23]) states that for a broad class of communication channel models, given $\delta > 0$ and $R$ lesser than the channel capacity, there exists an $[n;k]_q$ linear code with $\frac{k}{n} \geq R$ such that $P_e(C) < \delta$. In this section we state and prove a version of Shannon's theorem for valued information on a DSMC.

Let $C$ be an $[n;k]_q$ linear code. Given value functions $\mu_1, \mu_2 : C \to \mathbb{R}_+$ that differ by a constant, $\mu_1 = \lambda\mu_2$ for some $\lambda > 0$, the expected loss functions differ by the same constant hence we may say that $\mu_1$ and $\mu_2$ are *equivalent*. Let $[\mu]$ be the equivalence class of the value function $\mu$. The *canonical representative* of the class $[\mu]$ is defined to be the value function $\nu \in [\mu]$ such that $\|\nu\|_\infty = 1$, where $\|\nu\|_\infty$ denotes the maximum norm

$$\|\nu\|_\infty = \max\{\nu(c) : c \in C\}.$$

We can identify

$$[\mathcal{V}(C)] = \{[\mu] : \mu : C \to \mathbb{R}_+ \text{ value function}\},$$

the space of equivalence classes of value function, with the set of canonical representatives and consequently with the faces of the cube $[0,1]^{q^k}$. The value function $\mu_{\text{0-1}}$ corresponds to a vertex of $[0,1]^{q^k}$.

Let $\mathcal{V}_0(C)$ be the set of canonical representatives $\mu$ of of reasonable value functions on $C$ (i.e. $\mu(0) = 0$). Since $0 \leq \mu(c) \leq 1$ for all $c \in C$, it follows that:

$$\begin{aligned}
\mathbb{E}(a,\mu) &= \sum_{c\in C}\left(\sum_{y\in\mathbb{F}_q^n}\mu(a(y)-c)\,P(y|\,c)\right)P(c)\\
&= \sum_{c\in C}\left(\sum_{y\notin D(c)}\mu(a(y)-c)\,P(y|\,c)\right)P(c)\\
&\leq \sum_{c\in C}\left(1-\sum_{y\in D(c)}P(y|\,c)\right)P(c).
\end{aligned}$$

Thus $\mathbb{E}(a,\mu)$ is bounded by the decoding error probability:

$$\mathbb{E}(a,\mu) \leq P_e(C).$$

Therefore we have a version of Shannon's theorem for valued information on DSMC:

**Theorem 1** *For a DSMC let $\mathcal{C} > 0$ be the capacity of the channel. For each $\varepsilon > 0$ , $R < \mathcal{C}$ and $\mu \in \mathcal{V}_0$ there exists an $[n;k]_q$ linear code with $\frac{k}{n} \geq R$ such that $\mathbb{E}(a,\mu) < \varepsilon$.*

**Open Problem 1** *As was seen in the proof of the preceding theorem, since* $\mathbb{E}(a,\mu) \leq \mathbb{E}(a,\mu_{0-1}) = P_e(C)$*, we ask if it is possible to achieve reliable communication at rates superior to the Shannon capacity. In other words, for a given value function* $\mu$*, given* $\varepsilon > 0$ *there is a code* $C$ *such that* $\mathbb{E}_C(a,\mu) < \varepsilon$ *(Theorem 1). Let* $n_\mu(\varepsilon)$ *be the minimal possible length of such a code, so that the code has information rate* $\frac{k}{n_\mu(\varepsilon)}$ *(where* $k$ *is the dimension of the code)*[3]*. Since* $\mathbb{E}(a,\mu) \leq \mathbb{E}(a,\mu_{0-1})$ *we have that* $n_\mu(\varepsilon) \leq n_{\mu_{0-1}}(\varepsilon)$ *and we ask for a characterization of the value functions for which* $n_\mu(\varepsilon) < n_{\mu_{0-1}}(\varepsilon)$ *for every* $\varepsilon$*. Moreover, we ask if there is a value function* $\mu$ *such that*

$$\lim_{\varepsilon\to 0}\frac{n_\mu(\varepsilon)}{n_{\mu_{0-1}}(\varepsilon)} < 1$$

*or even*

$$\lim_{\varepsilon\to 0}\frac{n_\mu(\varepsilon)}{n_{\mu_{0-1}}(\varepsilon)} = 0.$$

# 6 Poset Metrics and Expected Loss Differences

The determination of Bayes decoders is a hard (in terms of complexity) problem. In order to have any hope to actually developing a communication process that needs, at its very end, a decoding algorithm, we shall consider a particular but large class of decoders, the nearest-neighbor (NN) decoders determined by poset metrics. Besides the fact of being a metric, those metrics profiteers well the structure of linear codes, since they are invariant by translations. As we shall explain latter, for many of those metrics there are very efficient decoding algorithms available.

Poset metrics were introduced in the context of coding theory by Richard Brualdi et. al. in 1995 (see [6]). Since its introduction in 1995 many contributions have been established for the theory of poset codes. The works on the existence of new classes of perfect codes (see [10], [12]), determination of identities of MacWilliams type ([1], [14]), Wei duality theorem ([17]), $P$-MDS codes ([11]) and the isometry groups ([18]) are examples of these contributions. Some particular families of poset metrics are also studied, the most common one is the family of Niederreiter-Rosenbloom-Tsfasman metrics ([3], [19], [20], [21]), since transmission over a set of parallel channels subject to fading and the noise process in a wireless fading system (see [21], [26]) are suitable to be modeled with such metrics.

[3] The decoder that is used to achieve such minimality is not relevant at this point, only the minimality of $n_\mu(\varepsilon)$.

We start defining what a poset metric is. Let $[n] := \{1, 2, \dots, n\}$ be a finite set with $n$ elements and let $\preceq$ be a *partial order* on $[n]$. We call the pair $P = ([n], \preceq_P)$ a *poset*. When no confusion may arise we will write simply $\preceq$ instead of $\preceq_P$. An *ideal* in $P$ is a subset $I$ satisfying the following condition: if $j \in I$ and $i \preceq j$, then $i \in I$. Given a subset $X$ in $P$, we denote by $\langle X \rangle$ the smallest ideal containing $X$, called the *ideal generated* by $X$. If $x = (x_1, \dots, x_n)$ and $y = (y_1, \dots, y_n)$ are two vectors in $\mathbb{F}_q^n$, then their *$P$-distance* $d_P(x, y)$ is defined by

$$d_P(x, y) = |\langle \{i : x_i \neq y_i\} \rangle|,$$

where $|A|$ denotes the cardinality of $A$. Since the $P$-distance is a metric on $\mathbb{F}_q^n$, it is also called *poset metric* (or *$P$-metric*). If $P$ is an *antichain order* (or *Hamming order*), that is, an order where $i \preceq j$ *iff* $i = j$, the $P$-distance is just the classical Hamming distance.

Before we move to look for expected loss for poset decoders, we introduce briefly two families of posets that will be considered along this work.

A *chain order* over $[n]$ is an an order where every two elements are comparable (see figure 5). A *Niederreiter-Rosenbloom-Tsfasman $(n, m)$-order* (NRT) over $[nm]$ is an order formed by the disjoint union of $n$ chains, each chain having $m$ elements (we call $m$ the *length* of the chain).

A *hierarchical order* $P$ over $[n]$ is an order for which there is a partition

$$[n] = \bigcup_{\delta=1}^{h} A_\delta$$

such that given $i \in A_{\delta_i}$ and $j \in A_{\delta_j}$, then $i \preceq_P j$ if, and only if, $\delta_i \leq \delta_j$. If we denote $|A_\delta| = l_\delta$ we may say $P$ is an $(l_1, \dots, l_h)$ *hierarchical poset* (see Figure 5). We remark that an $(1, \dots, 1)$ hierarchical poset is the $(1, n)$ NRT-order and an $(n)$ hierarchical poset is the chain order.

We stress that the first poset (with only trivial relations) gives rise to the usual Hamming metric $d_H$ and the second one may also be viewed as a $(1, 3)$ NRT order.

An ordinary decoder $a_P$ of an $[n; k]_q$ linear code $C$ is called an *nearest-neighbor $P$-decoder* ($P$-NN) if

$$d_P(y, a_P(y)) = \min_{c \in C} d_P(y, c)$$

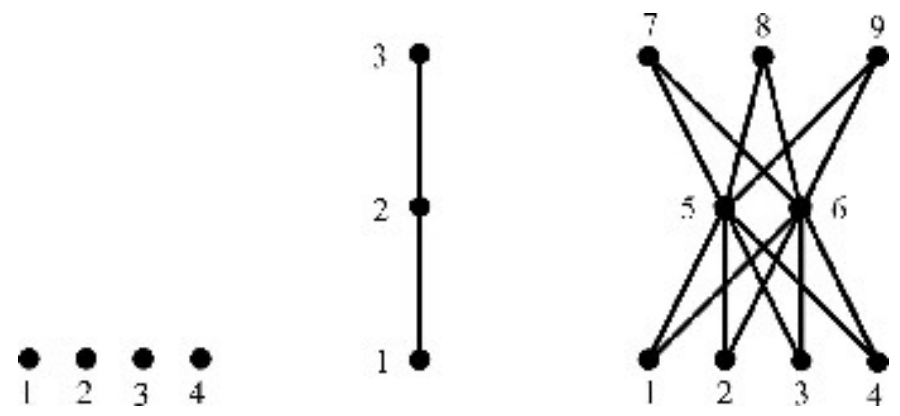

Figure 5: Hierarchical poset of type (4), (1, 1, 1) and (4, 2, 3).

for every $y \in \mathbb{F}_q^n$. The set of all $P$-NN decoders associated with $C$ will be denoted by $\mathcal{O}_P(C)$. We denote by $\mathcal{O}(C)$ the set of all such decoders, for all poset metrics $d_P$ in $\mathbb{F}_q^n$:

$$\mathcal{O}(C) := \bigcup_P \mathcal{O}_P(C) = \{a_P : a_P \text{ is a } P\text{-NN decoder of } C\}.$$

We remark that all those are reasonable decoders.

Decoders in $\mathcal{O}(C)$ are called *poset decoders.* A $Q$-NN decoder $a_Q^* \in \mathcal{O}(C)$ such that

$$\mathbb{E}\left(a_Q^*, \mu\right) = \min_{a_P \in \mathcal{O}(C)} \mathbb{E}(a_P, \mu)$$

is said to be a *Poset-Bayes decoder* for $C$ *relative to the value function* $\mu$. The determination of Poset-Bayes decoders is a hard problem, since the quantity of such decoders (for a fixed code $C$) grows exponentially with $n$. Our strategy is to consider the difference between the total expected loss relative of pairs of decoders in $\mathcal{O}(C)$.

Let $P = ([n], \leq)$ and $Q = ([n], \leq)$ be posets on $[n]$. Given a linear code $C \subseteq \mathbb{F}_q^n$, we consider two $P$-NN and $Q$-NN decoders $a_P, a_Q : \mathbb{F}_q^n \to C$ with total expected loss functions $\mathbb{E}_{a_P}(\mu) := \mathbb{E}(a_P, \mu)$ and $\mathbb{E}_{a_Q}(\mu) := \mathbb{E}(a_Q, \mu)$ respectively. The *total expected loss difference* between $\mathbb{E}_{a_P}(\mu)$ and $\mathbb{E}_{a_Q}(\mu)$ is

$$\mathbb{E}_{\left(a_P, a_Q\right)}(\mu) := \mathbb{E}_{a_P}(\mu) - \mathbb{E}_{a_Q}(\mu).$$

From Proposition 1 it follows that for each $0 < s < 1$ (we recall that $s$ depends only on the crossover probability $p$) the total expected loss difference can be viewed as the restriction to the positive octant $\mathbb{R}_+^M \subseteq \mathbb{R}^M$ ($M = q^k$) of the linear functional $\mathbb{E}_{\left(a_P,a_Q\right)} : \mathbb{R}^M \to \mathbb{R}$ given by

$$\mathbb{E}_{\left(a_P,a_Q\right)}(\mu) = \sum_{c\in C} T_{\left(a_P,a_Q,c\right)}(s)\,\mu(c)$$

where

$$T_{\left(a_P,a_Q,c\right)}(s) = G_{a_P}(c) - G_{a_Q}(c) := G_{(a_P,c)}(s) - G_{\left(a_Q,c\right)}(s)$$

and $G_{\left(a_{(\cdot)},c\right)}(s)$ is defined as in Proposition 1. Let us label the codewords as $C = \{c_0, c_1, \ldots, c_{M-1}\}$. Assuming that $\mathbb{E}_{a_P}(\mu) \neq \mathbb{E}_{a_Q}(\mu)$, $\mathbb{E}_{\left(a_P,a_Q\right)}$ is a non-null operator, then

$$\tau_{\left(a_P,a_Q\right)}(s) = \left(T_{\left(a_P,a_Q,c_0\right)}(s), \ldots, T_{\left(a_P,a_Q,c_{M-1}\right)}(s)\right)$$

is a vector in $\mathbb{R}^M$ orthogonal to $N_{\left(a_P,a_Q\right)}$, the kernel of $\mathbb{E}_{\left(a_P,a_Q\right)}$, and it points toward the connected component of $\mathbb{R}^M - N_{\left(a_P,a_Q\right)}$ containing those functions $\mu$ for which $\mathbb{E}_{\left(a_P,a_Q\right)}(\mu) > 0$.

If $\mathcal{V}$ denotes the set of all value functions (the positive octant of $\mathbb{R}^M$), it can be decomposed as

$$\mathcal{V} = \mathcal{V}^+_{\left(a_P,a_Q\right)} \cup \left(N_{\left(a_P,a_Q\right)} \cap \mathcal{V}\right) \cup \mathcal{V}^-_{\left(a_P,a_Q\right)}$$

where $\mathcal{V}^+_{\left(a_P,a_Q\right)}$ and $\mathcal{V}^-_{\left(a_P,a_Q\right)}$ are the subsets of value functions $\mu \in \mathcal{V}$ for which $\mathbb{E}_{\left(a_P,a_Q\right)}(\mu) > 0$ and $\mathbb{E}_{\left(a_P,a_Q\right)}(\mu) < 0$ respectively. We note that $\mathcal{V}^+_{\left(a_P,a_Q\right)}$ and $\mathcal{V}^-_{\left(a_P,a_Q\right)}$ are both non-empty iff $N_{\left(a_P,a_Q\right)}$ intersect the set of value functions $\mathcal{V}$, the positive octant of $\mathbb{R}^M$. Since each $\mu(c) \geq 0$, a necessary and sufficient condition for the kernel $N_{\left(a_P,a_Q\right)}$ intersecting the first quadrant of $\mathbb{R}^M$ is that at least two coordinates of the normal vector $\tau_{\left(a_P,a_Q\right)}(s)$ (the coefficients of the linear combination $\mathbb{E}_{\left(a_P,a_Q\right)}(\mu)$) have different signs.

With this notation we give the following natural definition:

**Definition 1** *Given an $[n;k]_q$ linear code $C$, a value function $\mu$ for $C$, and decoders $a_P, a_Q \in \mathcal{O}(C)$, we say that $a_P$ is* ***better*** *than $a_Q$ relative to $\mu$ if*

$$\mathbb{E}_{\left(a_P,a_Q\right)}(\mu) < 0,$$

*that is, if $\mathbb{E}_{a_P}(\mu) < \mathbb{E}_{a_Q}(\mu)$.*

We note that saying that *$a_P$ is better than $a_Q$ relative to $\mu$* is just a way of emphasizing the meaning of the statement $\mu \in \mathcal{V}^{+}_{\left(a_P,a_Q\right)}$. With the definition and notation above, we have actually proved the following:

**Theorem 2** *Let $C$ be an $[n;k]_q$ linear code. Given two poset decoders $a_P, a_Q \in \mathcal{O}(C)$, then there are value functions for which $a_P$ is better than $a_Q$ and value functions for which $a_Q$ is better than $a_P$ iff there are $c, c' \in C$ such that*

$$T_{\left(a_P,a_Q,c\right)}(s) < 0 < T_{\left(a_P,a_Q,c'\right)}(s).$$

*In this case, both $\mathcal{V}^{+}_{\left(a_P,a_Q\right)}$ and $\mathcal{V}^{-}_{\left(a_P,a_Q\right)}$ are nonempty subsets of $\mathcal{V}$.*

Bayes decoders associated to the value function $\mu_{\text{0-1}}$ are the classical ML decoders (see for example [2, Theorem 4.1.1]). In the context of expect loss and restricting the problem to the class of Poset-Bayes decoders, we have:

**Theorem 3** *Let $H$ be the Hamming order on $[n]$ and $C$ be an $[n;k]_q$ linear code. Then*

$$\mathbb{E}_{(a_H,a_P)}(\mu_{\text{0-1}}) \leq 0$$

*for any poset $P$ and any $a_P \in \mathcal{O}(C)$, that is, $H$-NN is better than $P$-NN for all order $P$ on $[n]$. Therefore, $H$-NN decoders are Poset-Bayes decoders for the value function $\mu_{\text{0-1}}$.*

Up to this point there is no real advantage in dealing with poset decoders. Such advantages will arise if we can give positive answer to the following questions:

(1) Given a linear code $C$ and the Hamming metric $d_H$, are there a $H$-NN decoder $a_H$ and a poset decoder $a_P$ such that $\mathcal{V}^{+}_{(a_H,a_P)}$ and $\mathcal{V}^{-}_{(a_H,a_P)}$ are nonempty? A positive answer would means that, for a nonempty set of value functions, the poset decoder $a_P$ is better than any $H$-NN decoder $a_H$.

(2) Given $P$ and $Q$ posets, are there a linear code $C$ and decoders $a_P$ and $a_Q$ of $C$ such that $\mathcal{V}^{+}_{(a_P,a_Q)}$ and $\mathcal{V}^{-}_{(a_P,a_Q)}$ are nonempty? A positive answer to this question means that every poset decoder is relevant, depending on the code under consideration.

Partial answers to those questions are given in Section 7.

The following examples illustrate the concepts and questions presented in this section.

**Example 1** *Let $H$ be the Hamming order and $P$ be the total order $1 \preceq_P 2 \preceq_P 3 \preceq_P 4 \preceq_P 5$. For the $[5;2]_2$ binary code*

$$C = \{c_0 = 00000, c_1 = 11100, c_2 = 00111, c_3 = 11011\}$$

*and appropriate decoders $a_H$ and $a_P$ of $C$ we have*

$$\begin{aligned}\mathbb{E}_{a_H}(\mu) = {} & \left(4 + 20s + 8s^2\right)\mu(c_0) + \\ & \left(12s^2 + 12s^3 + 8s^4\right)\mu(c_1) + \\ & \left(12s^2 + 12s^3 + 8s^4\right)\mu(c_2) + \\ & \left(8s^2 + 16s^3 + 4s^4 + 4s^5\right)\mu(c_3),\end{aligned}$$

$$\begin{aligned}\mathbb{E}_{a_P}(\mu) = {} & \left(4 + 10s + 10s^2 + 6s^3 + 4s^4\right)\mu(c_0) + \\ & \left(6s + 14s^2 + 10s^3 + 2s^4\right)\mu(c_1) + \\ & \left(2s + 6s^2 + 10s^3 + 10s^4 + 4s^5\right)\mu(c_2) + \\ & \left(2s + 10s^2 + 14s^3 + 6s^4\right)\mu(c_3),\end{aligned}$$

*hence*

$$\begin{aligned}\mathbb{E}_{(a_H,a_P)}(\mu) = {} & \left(10s - 2s^2 - 6s^3 + 2s^4\right)\mu(c_0) + \\ & \left(-6s - 2s^2 + 2s^3 + 6s^4\right)\mu(c_1) + \\ & \left(-2s + 6s^2 + 2s^3 - 2s^4 - 4s^5\right)\mu(c_2) + \\ & \left(-2s - 2s^2 + 2s^3 - 2s^4 + 4s^5\right)\mu(c_3).\end{aligned}$$

*Since for each $0 < s < 1$*

$$T_{(a_H,a_P,c_0)}(s) = 10s - 2s^2 - 6s^3 + 2s^4 > 0$$

*and*

$$T_{(a_H,a_P,c_1)}(s) = -6s - 2s^2 + 2s^3 + 6s^4 < 0,$$

*it follows from Theorem 2 that both* $\mathcal{V}^{+}_{(a_H,a_P)}$ *and* $\mathcal{V}^{-}_{(a_H,a_P)}$ *are nonempty subsets of* $\mathcal{V}$*, that is, depending on the value function, the* $P$*-NN decoder* $a_P$ *may be better or worse then the usual* $H$*-NN decoder* $a_H$*.*

**Example 2** *Let us now consider the repetition code* $\{c_0 = 000, c_1 = 111\}$*. Although trivial, this is an MDS perfect code. Considering the ML decoder* $a_H$ *and a poset decoder* $a_P$ *determined by the poset* $P$ *defined by the relations* $1 \preceq_P 2 \preceq_P 3$ *we find that*

$$\mathbb{E}_{(a_H,a_P)}(\mu) = \left(2s - s^2\right)\mu(000) + \left(-2s + s^2\right)\mu(111)$$

*hence*

$$\mathcal{V}^{+}_{(a_P,a_H)} = \{\mu : \mu(000) > \mu(111)\}$$

*and*

$$\mathcal{V}^{-}_{(a_P,a_H)} = \{\mu : \mu(000) < \mu(111)\}.$$

# 7 Relevance of Decoders and Codes

In this section we show that in quite general instances, every encoding and every decoder may be relevant, depending on the value functions to be considered. All proofs are postponed to Appendix 10.

We start with the result which shows that for any linear code and any ML decoder, there are always value functions for which is better to use an non-ML decoder.

**Theorem 4** *Let* $C$ *be an* $[n;k]_q$ *linear code and* $\widetilde{a}$ *an ML decoder. Than, there exist a decoder* $a$ *and value functions* $\mu$ *and* $\widetilde{\mu}$ *such that*

$$\mathbb{E}(a,\mu) > \mathbb{E}(\widetilde{a},\mu)$$

*and*

$$\mathbb{E}(a,\widetilde{\mu}) < \mathbb{E}(\widetilde{a},\widetilde{\mu}),$$

*for any given discrete channel.*

In the proof of this theorem we make use of a decoder $a_0 : \mathbb{F}_q^n \to C$ such that $a_0(y) = 0$ for all $y \in \mathbb{F}_q^n$ and of a value function $\mu_{\text{1-0}}$ defined by $\mu_{\text{1-0}}(0) = 1$ and $\mu_{\text{1-0}}(c) = 0$ if $c \neq 0$. Both this decoder and the value function are not reasonable ones.

We can make some progress concerning the question of reasonable decoders considering a discrete channel. Before proceeding, we given an $[n;k]_q$ linear code and $y \in \mathbb{F}_q^n$ we define $\arg(d_H(C,y))$ as the set of all codewords of $C$ closest to $y$ in the usual Hamming metric:

$$\arg(d_H(C,y)) = \left\{ c \in C : d_H(y,c) = \min_{\theta \in C} d_H(y,\theta) \right\}.$$

**Theorem 5** *Let $C$ be an $[n;k]_2$ binary linear code and $\widetilde{y} = (1,1,\ldots,1)$. If $|\arg(d_H(C,\widetilde{y}))| > 1$, there are ML decoders $a_H$ and $\widetilde{a}_H$ of $C$ such that $V^+_{(a_H,\widetilde{a}_H)}$ and $V^-_{(a_H,\widetilde{a}_H)}$ are both nonempty.*

Let us state a important class of linear codes that satisfies the condition of the Theorem 5.

**Corollary 1** *Let $d_H$ be the Hamming metric on $\mathbb{F}_2^n$ and $C$ be an $[n;k]_2$ binary linear code of constant weight $w$. If $k > 1$, then there is a $P$-NN decoder $a_P$ and an $H$-NN decoder $a_H$ of $C$ such that $\mathcal{V}^+_{(a_H,a_P)}$ and $\mathcal{V}^-_{(a_H,a_P)}$ are nonempty.*

**Open Problem 2** *Does Theorem 4 still hold if we impose the use of reasonable decoders and value functions? In Theorem 5 we considered a reasonable decoder (every P-NN decoder is reasonable), but had to impose some restrictions on the code. Is it possible to rule those conditions out? We do believe the answer to those questions is positive, but were not able to prove it.*

Up to this moment we were considering a given (and fixed) code, and showed there are value functions for which it is better (in the sense of minimizing the expected loss) to decode using a non-Hamming decoder. Now we fix two different posets, one of them a poset $P$ that satisfies a special condition and the usual Hamming poset $H$ and show the existence of a code for which both $\mathcal{V}^+_{(a_H,a_P)}$ and $\mathcal{V}^-_{(a_H,a_P)}$ are non empty.

Before stating the results we need some definitions. Given an order $P = ([n], \leq_P)$ the *dual order* $P^* = ([n], \leq_{P^*})$ is defined by the opposite relations: $x \leq_{P^*} y \Leftrightarrow y \leq_P x$. For simplicity, we shall omit the indices in $\leq_P$ and $\leq_{P^*}$ when no confusion may be caused. We remark that $(P^*)^* = P$. An ideal in $P^*$ is called a *filter* in $P$.

Given a nontrivial and proper filter $I$ in $P$ and $\emptyset \neq J \subset I$, we define

$$I_J^+ := \{i \in I - J : i > j \text{ for some } j \in J\}$$

and

$$I_J^- := \{i \in I - J : i < j \text{ for some } j \in J\}.$$

We will say that a filter $I$ of $P$ is *$J$-decomposable* if

$$I = I_J^+ \cup J \cup I_J^-$$

is a partition of $I$ with both $I_J^+$ and $I_J^-$ nonempty. If there exists a filter $I$ in $P$ that is $J$-decomposable, we will say that $P$ is *$(I, J)$-decomposable.*

Let now $\{e_i : 1 \leq i \leq n\}$ be the usual base of $\mathbb{F}_q^n$. For each nonempty subset $X \subseteq [n]$ let

$$C_X = \operatorname{span}\{e_i : i \in X\}$$

be the *coordinate subspace with support in* $X$. Given $y = \sum_{i=1}^n y_i e_i \in \mathbb{F}_q^n$ we denote by $y_X$ its projection onto $C_X$:

$$y_X = \sum_{i \in X} y_i e_i.$$

Given an $(I, J)$-decomposable order $P$, with $|I| = K$ and $|J| = k$, it determines an $[n; K - k]_q$ linear code $C_{(I,J)}$ that is just the coordinate space $C_{I-J}$, i.e.,

$$C_{(I,J)} = \operatorname{span}\{e_i : i \in I - J\}.$$

We name those subspaces as a *BGL code*, after the description of perfect codes given by Brualdi, Graves and Lawrence in 1995 ([6, Theorem 2.1]).

The *complement* of a subset $X \subset [n]$ is denoted by $X^c$.

**Theorem 6** *Let $P = ([n], \leq_P)$ be an $(I, J)$-decomposable order and $H = ([n], \leq_H)$ be the Hamming order. Considering the $[n; |I| - |J|]_q$ BGL code $C_{(I,J)}$, there are NN decoders $a_H$ and $a_P$ of $C$ and codewords $c, c' \in C_{(I,J)}$ such that both $\mathcal{V}^+_{(a_H,a_P)}$ and $\mathcal{V}^-_{(a_H,a_P)}$ are non empty for every $0 < s < 1$.*

In general is not easy to compute the polynomial $T_{(a_H,a_P,c)}(s)$. However, for appropriate $P$-NN decoders of $C_{(I,J)}$ it is possible to determine $T_{(a_H,a_P,c)}(s)$:

**Corollary 2** *Consider an* $(I, J)$*-decomposable order* $P$ *on* $[n]$*. For the BGL code* $C_{(I,J)}$ *and for the* $P$*-NN decoder* $a_P$ *determined in Theorem 6 we have that*

$$T_{(a_H, a_P, c)}(s) = s^{d_H(\widetilde{y}, c)} - s^{d_H(\widetilde{y}, \widetilde{c} - c)}$$

*for every* $c \in C_{(I,J)}$.

It is easy to see that the class of $(I, J)$-decomposable orders includes the $(n, m)$-NRT poset for $m \geq 4$, hence the following corollary holds:

**Corollary 3** *Let* $H$ *be the Hamming order on* $[nm]$ *and let* $P$ *be the NRT* $(n, m)$*-order on* $[nm]$*.Then, for* $m \geq 4$ *there exists an* $[n; k]_q$ *linear code* $C$ *and* $c, c' \in C$ *such that*

$$T_{(a_H, a_P, c)}(s) < 0 < T_{(a_H, a_P, c')}(s)$$

*for some* $H$*-NN and* $P$*-NN decoders* $a_H$ *and* $a_P$ *of* $C$ *respectively. Therefore,* $\mathcal{V}^+_{(a_H, a_P)}$ *and* $\mathcal{V}^-_{(a_H, a_P)}$ *are nonempty.*

If we consider the particular case when $P$ is the $(1, m)$-NRT order and $m \geq 4$, then every filter $I$ of $J$ with $|I| \geq 3$ is decomposable: given $I = \{m - k + 1, \ldots, m\}$, then

$$I = I^+_{\{m-k+j\}} \cup \{m - k + j\} \cup I^-_{\{m-k+j\}}$$

is a non trivial partition of $I$ and the following holds:

**Corollary 4** *Let* $H$ *be the Hamming order on* $[m]$ *and let* $P$ *be the NRT* $(1, m)$*-order. If* $m \geq 4$*, then for each* $2 \leq k < m - 2$ *there exists an* $[m; k]_q$ *linear code* $C$ *such that both* $\mathcal{V}^+_{(a_H, a_P)}$ *and* $\mathcal{V}^-_{(a_H, a_P)}$ *are nonempty for some* $H$*-NN and* $P$*-NN decoders* $a_H$ *and* $a_P$.

We remark that, in the proof of Theorem 6, the only property of the Hamming poset $H$ we used was the fact that $a_H(y) = y_{I-J}$ is an $H$-NN decoder. Actually it is true for any poset on $[n]$ that is $(I, J)$-decomposable. It follows that the result obtained in Theorem 6 also holds for any such pair of posets. Let $P$ and $Q$ be a pair of orders on $[n]$. Suppose that $P$ is $(I, J)$-decomposable. We will say that $P$ is $(I, J)$*-isomorphic* to $Q$ if $I$ in $Q$ is still a filter. In this condition $I$ is also $J$-decomposable on $Q$.

**Theorem 7** *Consider proper subsets* $\emptyset \neq J \subset I \subset [n]$. *Let* $P$ *and* $Q$ *be posets on* $[n]$ *such that* $I$ *is filter in both* $P$ *and* $Q$. *Then, if both* $P$ *and* $Q$ *are* $(I, J)$*-decomposable there is an* $[n;k]_q$ *linear code* $C$ *and NN decoders* $a_P$ *and* $a_Q$ *of* $C$ *such that* $\mathcal{V}^{+}_{(a_P,a_Q)}$ *and* $\mathcal{V}^{-}_{(a_P,a_Q)}$ *are nonempty*[4].

**Open Problem 3** *We do believe that the* $(I, J)$*-decomposable condition in the statement of Theorem 6 is not necessary, as much as the condition that the channel being symmetric. The right question we believe should be posed is the following: to find necessary and sufficient conditions on two posets that guarantee the existence of a code that may be better corrected by using either the poset metrics (depending on the value functions).*

# 8 Value Functions for the Continuous Channel

The concept of expected loss defined for a discrete channel can naturally be adapted for continuous channels. We do not go as further as in the discrete channel case and restrict ourselves to giving appropriate definitions.

Let $S = \{s_1, \ldots, s_M\}$ be a finite *signal constellation* on the Euclidean $N$-dimensional space $\mathbb{R}^N$. We should now proceed to introduce value to the signal. In a manner of fact, in a situation similar to that developed for the discrete channel, we are actually valuing the errors (after decoding), what was not totally evident in the discrete case since we considered just linear codes, hence every error (after decoding) is a codeword. For this reason we consider the *difference set*

$$\Delta S := S \; - S = \{s_i - s_j : s_i, s_j \in S\}.$$

A *value function for the constellation* $S$ is any function

$$\mu : \Delta S \to \mathbb{R}_+.$$

Consider the *continuous channel* defined by the family of probability density functions $p(y|\, x)$, with $x, y \in \mathbb{R}^N$. We define the *overall expected loss* of

[4]BGL codes have some nice properties, including the possibility of expressing the packing radius as a function of the minimal distance, what is not an easy task for general codes and posets.

$S$ relative to the value function $\mu : \Delta S \to \mathbb{R}_+$ and decoder $a : \mathbb{R}^N \to S$ as the average

$$\mathbb{E}\left(\mathcal{L}\left(a,\mu\right)\right) = \int_{\mathbb{R}^n} \mathcal{L}_y\left(a,\mu\right) p\left(y\right) dy$$

where

$$\mathcal{L}_y\left(a,\mu\right) = \sum_{s_i \in S} \mu\left(a\left(y\right) - s_i\right) p\left(s_i | \, y\right) dy$$

is the *expected loss for an observed $y$*. As in the discrete channel case, $\mathbb{E}\left(a, \mu_{0\text{-}1}\right)$ coincides with the decoding error probability $P_e\left(S\right)$ of $S$. As in the discrete case, we denote $\mathbb{E}\left(\mathcal{L}\left(a,\mu\right)\right)$ simply by $\mathbb{E}\left(a,\mu\right)$

Also the overall expected loss $\mathbb{E}\left(a,\mu\right)$ can be interpreted as the restriction of a linear functional with domain $\mathbb{R}^{|\Delta S|}$ into $\mathbb{R}_+$:

$$\mathbb{E}\left(a,\mu\right) = \sum_{\tau \in \Delta S} G_a\left(\tau\right) \mu\left(\tau\right)$$

with

$$G_a\left(\tau\right) = \sum_{s_i, s_j \in S : s_j - s_i = \tau} \int_{R(s_j)} p\left(y | \, s_i\right) p\left(s_i\right) dy$$

where $R\left(s_i\right) = a^{-1}\left(s_i\right)$ is the decision region of the signal $s_i$.

Now consider the *difference of the expected losses*

$$\mathbb{E}_{(a,\widetilde{a})}\left(\mu\right) = \mathbb{E}\left(a,\mu\right) - \mathbb{E}\left(\widetilde{a},\mu\right)$$

relative to the value function $\mu$ and the pair $\left(a, \widetilde{a}\right)$ of decoders of $S$, we have that

$$\mathbb{E}_{(a,\widetilde{a})}\left(\mu\right) = \sum_{\tau \in \Delta S} T_{(a,\widetilde{a})}\left(\tau\right) \mu\left(\tau\right)$$

where

$$T_{(a,\widetilde{a})}\left(\tau\right) := G_a\left(\tau\right) - G_{\widetilde{a}}\left(\tau\right)$$

for each $\tau \in \Delta S$.

With the definitions properly established, it is possible to prove that ML decoders on $\mathbb{R}^N$, determined by the Voronoi regions, are not always the best decoders. More generally:

**Theorem 8** *Let* $S = \{s_1, \ldots, s_M\}$ *be a signal constellation in* $\mathbb{R}^N$ *such that for some* $\tau \in \mathbb{R}^N$ *there is a unique* $s_j - s_i \in \Delta S$ *such that* $s_j - s_i = \tau$*. Consider a decoder* $a : \mathbb{R}^N \to S$ *for* $S$ *and assume that each decision region of the decoder* $a$ *has non-empty interior. Then there is another decoder* $\widetilde{a} : \mathbb{R}^N \to S$ *for* $S$ *such that*

$$T_{(a,\widetilde{a})}(s_i - s_j) < 0 < T_{(a,\widetilde{a})}(s_j - s_i).$$

# 9 Final Remarks

In this final section we present some remarks concerning many aspects of coding theory that either demand a different formulation in the context of expected loss or raise interesting problems we believe are deserve to be explored.

## 9.1 Remarks about Poset Decoders

Poset codes were given a distinctive position in this work, but its actual importance was not truly explained. The first motivation to consider poset decoders is the fact that some of them admit efficient algorithm decoding, what is a deep contrast with the usual setting of ML decoders case, where finding general decoding algorithms is known to be NP-complete (see [4]). Indeed, for an $(n)$-hierarchical poset (or equivalently, an NRT $(1, n)$-order), the kind used to produce the right-side pictures in the Introduction, decoding algorithm is linear in the co-dimension of the code [20, Section IV-D]. Besides those posets, for a general hierarchical poset, there are algorithms that are at least as efficient as syndrome decoding and with high probability significantly faster [8]. If an $(n)$-hierarchical poset is unique (up to order isomorphism) for any $n \in \mathbb{N}$, the hierarchical posets in their generality are a large family, corresponding to ordered partitions of $n$, hence there are $\sim 2^{\sqrt{n}}$ such posets, what should provide many possibilities in each code length.

When considering the dimension $k$ and the length $n$ of a code as given, the usual task of error correcting is to find a code with better performance. If this is already an untractable computational problem, the goal posed in this work is much more complex: finding a pair, consisting of a code and a decoder. Here comes another reason to restrict ourselves to poset decoders, or more specifically, to hierarchical poset decoders, since there is an heuristic approach to find better results for the expected loss function: since coordinates that has large value (it means, $|\langle i \rangle|$ is large) are best protected, we should make a code such that the more relevant information is concentrated as non-null entries in those coordinates and to define a poset that has those coordinates as maximal elements of the poset.

**Open Problem 4** *Is it possible to prove that under suitable conditions this kind of heuristics will work? To be more explicit. Suppose there are $M = q^k$ informations $\Lambda = \{x_1, \ldots, x_M\}$ with $\mu(x_i) \sim \lambda^i$ for some constant $\lambda > 1$. Is it true that given an $[n,k]_q$ linear code $C$ it is possible to find an $a_P$ decoder determined by an NRT $(1,n)$-order $P$ for which $\mathbb{E}(a_P, \mu) < \mathbb{E}(a_H, \mu)$ for every Hamming decoder $a_H$? Can we make the same statement when $C$ is a perfect or MDS code? More generally, if the information set can be partitioned as*

$$\Lambda = \bigcup_{i=1}^{r} \Lambda_i$$

*with $\mu(x_j) \sim \lambda^i$ if $x_j \in \Lambda_i$, should we use a decoder determined by an hierarchical poset?*

## 9.2 Remarks about the Space of Codes and Decoders

The introduction of value functions and the need to consider decoders that may not be ML decoders enlarges considerably the space where we are actually working. We can assume as reasonable that the quantity $q^k$ of information and the value of the information are given, depending on the application and the kind of knowledge the information constitute. If we suppose for instance that the cost of transmission is determined by the length $n$ (the same would hold if a maximal bound was established for the expected loss), we are looking for a pair consisting of an $[n;k]_q$ linear code and a decoder associated to the code. Moreover, we are not only interested in the code $C \subseteq \mathbb{F}_q^n$, but actually how the set of information $\Lambda$ is mapped onto $C$, so we are actually considering a code not only as a subset $C \subseteq \mathbb{F}_q^n$ but as an embedding in $\mathbb{F}_q^n$. In other words, we should consider not only the subset $C \subseteq \mathbb{F}_q^n$ but also all the permutations $\sigma : C \to C$. In this sense, the pair $(\text{code}, \text{decoder})$ where we are searching for possibilities to minimize the expected loss function is a space with

$$\prod_{k=1}^{n} \frac{(q^n - q^{i-1})}{(q^k - q^{i-1})} \times \left(q^k\right)! \times (q^n)^{\left(2^k\right)}$$

elements, where the first factor corresponds to the cardinality of the Grasmannian $G(n,k)$, the second to the permutations of total quantity of $C$ and the last one to the decoders of a given $[n;k]_q$ code (including the unreasonable ones).

**Open Problem 5** *To estimate asymptotically, the quotient*

$$\frac{(q^n)^{(2^k)}}{DP_n}$$

*where $DP_n$ is the number of NN poset decoders may be an interesting question for itself. There is no known estimative of $DP_n$ but for $P_n$, the number of posets on a set of $n$ elements, the exact asymptotic is known [25]: for odd $n$*

$$P_n \sim \sqrt{\frac{2}{\pi}} \left( \sum_{x=-\infty}^{+\infty} 2^{-x^2} \right) 2^{\frac{n^2}{4}+\frac{3}{2}n-\frac{1}{2}\log n}$$

*and for even $n$*

$$P_n \sim \sqrt{\frac{2}{\pi}} \left( \sum_{x=-\infty}^{+\infty} 2^{-\left(x^2-\frac{1}{2}\right)^2} \right) 2^{\frac{n^2}{4}+\frac{3}{2}n-\frac{1}{2}\log n}.$$

## 9.3 Remarks about Bounds for Expected Loss

The error probability function $P_e(C)$ is one of the fundamental parameters to measure de performance of a coding scheme. Despite the fact it has a simple formulation,

$$P_e(C) = \frac{1}{M} \sum_{c \in C} \sum_{y \in D^c(c)} P(y|c),$$

actual calculations are generally hard problems. For this reason, finding good (lower and upper) bounds is a fundamental question. Among the well known such bounds we can find union bound, Bhattacharyya, Gallager, Caen and sphere packing (see [22]).

Considering the total expected loss, we already saw that the error probability $P_e(C)$ is an upper bound for $\mathbb{E}(a,\mu)$, but it is obviously far from being a tight one. Calculating $\mathbb{E}(a,\mu)$ it is not only prohibitive, but involves many parameters that should be treated separately.

**Open Problem 6** *Consider a fixed family of value functions and search for upper bounds for* $\mathbb{E}(a,\mu)$. *One relevant family that may be interesting for protecting information of different nature (as done in [5]) suggests to consider a value function* $\mu_{0\text{-}\lambda\text{-}1}$ *such that* $C$ *can be partitioned as* $C = C_0 \cup C_\lambda \cup C_1$ *where* $\mu_{0\text{-}\lambda\text{-}1}(c) = j$ *if* $c \in C_j$. *More general situations are found where* $C$ *is expressed as*

$$C = C_0 \cup C_1 \cup \ldots \cup C_r$$

*and the value function is either linear* ($\mu_L$) *or exponential* ($\mu_E$), *in the sense that*

$$\mu_L(c) = j\frac{1}{r} \text{ if } c \in C_j$$

$$\mu_E(c) = b^j \frac{1}{b^r} \text{ if } c \in C_j,$$

*where* $b > 1$ *is a constant. We remark that the fractions* $\frac{1}{r}$ *and* $\frac{1}{b^r}$ *are just scaling factors.*

**Open Problem 7** *Considering a family of value functions concerns aiming to produce coding schemes for families of applications with similar semantic value, and this is a data that is determined by the practical (or theoretical) problem. If instead we look at the way we are able to manage, the natural question would be to find bounds for the expected loss when considering a particular family of NN poset decoders, specially those determined by hierarchical posets.*

Finally:

**Open Problem 8** *Both the problems presented above are still very hard, in each of them there is one free parameter we do not find in the classical case where both the value function* ($\mu_{0\text{-}1}$) *and the decoder type (ML) are fixed. So we can combine the two previous problems and ask to find bounds for the expected loss function fixing a family of value functions and a type of NN decoder.*

## 9.4 Remarks about Rate Distortion Theory

The basic question concerning rate distortion theory, as posed Kolmogorov ([15]) and Shannon ([24]) is: given a source distribution and a distortion measure, what is the minimum rate description required to achieve a particular distortion? For expected loss function, the question may be stated as follows:

**Open Problem 9** *Let $\lfloor x \rfloor$ denote the integer part of $x \in \mathbb{R}$. Let $\mathcal{I}$ be a set of information and $\mu$ a value function defined on $\mathcal{I}$. Given a loss $E$, what is the maximal information rate $R$ for which there is an $\left[\left\lfloor \frac{k}{R} \right\rfloor; k\right]_q$ linear code $C$ and a $P$-NN decoder $a_P$ of $C$ such that*

$$\mathbb{E}\left(a_P, \mu\right) \leq E?$$

The basic definitions of rate distortion theory (see [7]) can be re-stated in the context of value functions and expected loss.

**Definition 2** *Let $I = \{x_1, \ldots, x_k\}$ be an information set and $\mu$ a value function on $I$. We say that the rate loss pair $(R, E)$ is* ***realizable*** *if there is an $\left[\left\lfloor \frac{k}{R} \right\rfloor; k\right]_q$ linear code $C$ and a $P$-NN decoder $a_P$ such that $\mathbb{E}\left(a_P, \mu\right) \leq E$. The* ***rate loss region*** *of $I$ is the closure of all realizable rate loss pairs. The* ***rate loss function*** *$R(E)$ is the maximum $R$ such that $(R, E)$ is in the rate loss region of $I$. The* ***capacity*** *$\mathcal{C}_\mu$ of the channel to transmit information from $\mathcal{I}$ given the value function $\mu$ is*

$$\mathcal{C}_\mu = \lim_{E \to 0} R(E).$$

**Open Problem 10** *Determine $\mathcal{C}_\mu$ for a family of value functions, restricted to a family of poset decoders.*

# 10 Appendix 1: Proofs

## 10.1 Proof of Theorem 4

Let $C$ be an $[n; k]_q$ linear code and $\mu_{\text{1-0}} : C \to \mathbb{R}_+$ be the value function such that $\mu_{\text{1-0}}(0) \neq 0$ and $\mu_{\text{1-0}}(c) = 0$ for all $0 \neq c \in C$. Let us consider the decoder $a_0 : \mathbb{F}_q^n \to C$ such that $a_0(y) = 0$ for all $y \in \mathbb{F}_q^n$. The total expected loss $\mathbb{E}\left(a_0, \mu_{\text{1-0}}\right)$ may be determined without utilizing the expressions in Section 4.

When a codeword $c \in C$ is transmitted, a word $y_c$ is received and it is decoded as $a_0(y_c)$. We remark that this decoding results in a loss $\mu_{\text{1-0}}(a_0(y_c) - c)$. However, $a_0(y_c) = 0$ for every $y_c$ hence the loss is just $\mu_{\text{1-0}}(0 - c) = \mu_{\text{1-0}}(-c)$. But $\mu_{\text{1-0}}(-c) \neq 0$ iff $c = 0$ and since we are assuming codewords are to be send with probability equal to $\frac{1}{q^k}$ we find that

$$\mathbb{E}\left(a_0, \mu_{1\text{-}0}\right) = \frac{\mu_{1\text{-}0}\left(0\right)}{q^k}$$

and this does not depends on the channel.

Given a decoder $a$ we have that

$$\begin{aligned}\mathbb{E}\left(a, \mu_{1\text{-}0}\right) &= \sum_{\tau \in C} G_a\left(\tau\right) \mu_{1\text{-}0}\left(\tau\right) \\ &= G_a\left(0\right) \mu_{1\text{-}0}\left(0\right).\end{aligned}$$

Considering a discrete channel determined by the set of conditional probabilities $P\left(y|\, x\right)$ we have (as in expression (2)) that

$$G_a\left(\tau\right) = \sum_{y \in \mathbb{F}_q^n} P\left(y|\, a\left(y\right) - \tau\right) P\left(a\left(y\right)\right)$$

hence

$$G_a\left(0\right) = \sum_{y \in \mathbb{F}_q^n} P\left(y|\, a\left(y\right)\right) P\left(a\left(y\right)\right).$$

Assuming that the probability distribution $P\left(c\right)$ of $C$ is uniform, we find that

$$G_a\left(0\right) = \frac{1}{q^k} \sum_{y \in \mathbb{F}_q^n} P\left(y|\, a\left(y\right)\right).$$

Considering an ML decoder $\tilde{a}$, we have by definition of ML decoder that

$$P\left(y|\, \widetilde{a}\left(y\right)\right) \geq P\left(y|\, c\right)$$

for every $c \in C$ so that

$$\sum_{y \in \mathbb{F}_q^n} P\left(y|\, \widetilde{a}\left(y\right)\right) \geq \sum_{y \in \mathbb{F}_q^n} P\left(y|\, c\right) = 1 \tag{3}$$

for every $c \in C$. Since for $c \in C$ and $y \notin \widetilde{a}^{-1}\left(c\right)$ we have that

$$P\left(y|\, \widetilde{a}\left(y\right)\right) > P\left(y|\, c\right),$$

we find that inequality (3) is actually strict:

$$\sum_{y\in\mathbb{F}_q^n} P\left(y|\,\widetilde{a}\left(y\right)\right) > 1. \tag{4}$$

So, since

$$\begin{aligned}\mathbb{E}\left(\widetilde{a},\mu_{1\text{-}0}\right) &= \sum_{\tau\in C} G_{\widetilde{a}}\left(\tau\right)\mu_{1\text{-}0}\left(\tau\right)\\ &= G_{\widetilde{a}}\left(0\right)\mu_{1\text{-}0}\left(0\right)\\ &= \frac{1}{q^k}\sum_{y\in\mathbb{F}_q^n} P\left(y|\,\widetilde{a}\left(y\right)\right)\mu_{1\text{-}0}\left(0\right), \end{aligned} \tag{5}$$

by (4) and (5) we have that

$$\mathbb{E}\left(\widetilde{a},\mu_{1\text{-}0}\right) > \frac{\mu_{1\text{-}0}\left(0\right)}{q^k}$$

and since $\mathbb{E}\left(a_0,\mu_{1\text{-}0}\right) = \frac{\mu_{1\text{-}0}(0)}{q^k}$ we conclude that

$$\mathbb{E}\left(\widetilde{a},\mu_{1\text{-}0}\right) > \mathbb{E}\left(a_0,\mu_{1\text{-}0}\right).$$

To finish we just consider the decoder $\tilde{a} = a_0$ above defined, $\mu = \mu_{0\text{-}1}$ and $\tilde{\mu} = \mu_{1\text{-}0}$.

## 10.2 Proof of Theorem 5

Let $d_H$ be the usual Hamming metric and let $c_1, c_2 \in \arg\left(d_H\left(C,\widetilde{y}\right)\right)$ with $c_1 \neq c_2$. Define $a_H\left(\widetilde{y}\right) = c_1$ and let $\widetilde{a}_H$ be defined by

$$\widetilde{a}_H\left(y\right) = \begin{cases} a_H\left(y\right) & \text{if} \quad y \neq \widetilde{y} \\ c_2 & \text{if} \quad y = \widetilde{y} \end{cases}.$$

With this definitions of $a_H$ and $\tilde{a}_H$ we find that

$$d_H\left(\widetilde{y}, a_H\left(\widetilde{y}\right) - c_1\right) = n,$$

$$d_H\left(\widetilde{y}, \widetilde{a}_H\left(\widetilde{y}\right) - c_1\right) = d_H\left(\widetilde{y}, c_2 - c_1\right),$$

$$d_H\left(\widetilde{y}, a_H\left(\widetilde{y}\right) - c_2\right) = d_H\left(\widetilde{y}, c_1 - c_2\right)$$

and

$$d_H\left(\widetilde{y}, \widetilde{a}_H\left(\widetilde{y}\right) - c_2\right) = n.$$

Setting $m = d_H\left(\widetilde{y}, c_2 - c_1\right) = d_H\left(\widetilde{y}, c_1 - c_2\right)$ we get

$$T_{(a_H, \widetilde{a}_H)}\left(c_1\right) = s^n - s^m$$

and

$$T_{(a_H, \widetilde{a}_H)}\left(c_2\right) = s^m - s^n.$$

Since $c_1 \neq c_2$, we have that $m < n$ and hence

$$T_{(a_H, \widetilde{a}_H)}\left(c_1\right) < 0 < T_{(a_H, \widetilde{a}_H)}\left(c_2\right)$$

for every $0 < s < 1$.

## 10.3 Proof of Corollary 1

Let $\widetilde{y} = (1, 1, \ldots, 1)$. Since $k > 1$ and $C$ has constant weight we have that $\widetilde{y} \notin C$. If $0 \neq c \in C$ is a codeword, since $C$ has constant weight $w$ it follows that

$$d_H\left(\widetilde{y}, c\right) = n - w_H\left(c\right) = n - w$$

and consequently $\arg\left(d_H\left(C, \widetilde{y}\right)\right) = C - \{0\}$ and $\left|\arg\left(d_H\left(C, \widetilde{y}\right)\right)\right| = 2^k - 1$. Since we are assuming $k > 1$, we conclude that $\left|\arg\left(d_H\left(C, \widetilde{y}\right)\right)\right| > 1$ and the result follows from Theorem 5.

## 10.4 Proof of Theorem 6

In this proof the complement of a subset $X$ of $[n]$ is denoted by $X^c$.

A vector $y \in \mathbb{F}_q^n$ can be decomposed as

$$y = y_{I^c} + y_J + y_{I-J}$$

where $y_{I^c}$, $y_J$ and $y_{I-J}$ are the projections of $y$ in the coordinate subspaces $C_{I^c}$, $C_J$ and $C_{(I,J)}$ respectively. From this decomposition follows that

$$d_H\left(y, c\right) = w_H\left(y_{I^c}\right) + w_H\left(y_J\right) + d_H\left(y_{I-J}, c\right)$$

for every $c \in C_{(I,J)}$. It follows that

$$\begin{aligned}
\arg \min_{\theta \in C_{(I,J)}} d_H(y,\theta) &= \arg \min_{\theta \in C_{(I,J)}} \{w_H(y_{I^c}) + w_H(y_J) + d_H(y_{I-J},\theta)\} \\
&= \arg \min_{\theta \in C_{(I,J)}} d_H(y_{I-J},\theta) \\
&= y_{I-J},
\end{aligned}$$

hence $a_H(y) = y_{I-J}$.

We claim that

$$a_P(y) = y_{I-J}$$

for every $y \in C_{J^c}$, that is, for such an $y$ we have that $y_J = 0$. Indeed, this happens because

$$d_P(y,c) = \begin{cases} |\langle supp(y_{I^c}) \cup supp(y_{I-J} - c)\rangle| & \text{if} \quad c \neq y_{I-J} \\ |\langle supp(y_{I^c})\rangle| & \text{if} \quad c = y_{I-J} \end{cases}.$$

It follows that

$$a_P(y) = \arg \min_{\theta \in C_{(I,J)}} d_P(y,\theta) = y_{I-J},$$

hence

$$a_P(y) = a_H(y)$$

for $y \in C_{J^c}$.

We should now define a $P$-NN decoder for $y \notin C_{J^c}$. So, we consider $y = y_{I^c} + y_J + y_{I-J}$ with $y_J \neq 0$. If $J' = supp(y_J)$ and $y_{I_J^+}$ is the projection of $y$ on $C_{I_J^+}$, then

$$\widetilde{y}_{I_{J'}^-} + y_{I_J^+} = \arg \min_{\theta \in C_{(I,J)}} d_P(y,\theta)$$

for every $\widetilde{y}_{I_{J'}^-} \in C_{I_{J'}^-}$. At this point we should note that the $H$-NN decoder $a_H$ already defined is also a $P$-NN decoder, since $y_{I-J} = \widetilde{y}_{I_{J'}^-} + y_{I_J^+}$ for some $\widetilde{y}_{I_{J'}^-} \in C_{I_{J'}^-}$. But this will not serve to our purpose, since in this case the total expected loss difference is 0. However, there are other possibilities for a $P$-NN decoder and we will define $a_P$ in such a way that

$$T_{(a_H,a_P,c')}(s) < 0 < T_{(a_H,a_P,c')}(s) \tag{6}$$

for some pair $c, c' \in C_{(I,J)}$.

Let

$$\widetilde{y} = \widetilde{x}_{I^c} + e_J$$

with $\widetilde{x}_{I^c} \in C_{I^c}$ and $e_J \in C_J$. Consider

$$\widetilde{c} \in C_{I_J^-}$$

with $\widetilde{c} \neq 0$. In this situation $\langle supp\,(\widetilde{y} - \widetilde{c})\rangle = \langle supp\,(\widetilde{y})\rangle$. Thus $a_P(\widetilde{y}) := \widetilde{c}$ is a $P$-NN decoder for $\widetilde{y}$. We conclude defining $a_P(y) = y_{I-J}$ for $\widetilde{y} \neq y \notin C_{J^c}$.

We now choose vectors $c_1, c_2 \in C_{(I,J)}$ that will ensure condition (6). We define

$$c_1 = \sum_{i \in I_J^+} e_i \in C_{(I,J)}$$

and

$$c_2 = \widetilde{c}.$$

Since $a_H(\widetilde{y}) = 0$ and $a_P(\widetilde{y}) = \widetilde{c}$, we have:

$$n_1 := d_H(\widetilde{y}, a_H(\widetilde{y}) - c_1) = d_H(\widetilde{y}, 0 - c_1) = w_H(\widetilde{y}) + \left|I_J^+\right|$$

and

$$m_1 := d_H(\widetilde{y}, a_P(\widetilde{y}) - c_1) = d_H(\widetilde{y}, \widetilde{c} - c_1) = w_H(\widetilde{y}) + w_H(\widetilde{c}) + \left|I_J^+\right|,$$

hence $n_1 < m_1$.

Moreover,

$$n_2 := d_H(\widetilde{y}, a_H(\widetilde{y}) - c_2) = d_H(\widetilde{y}, 0 - c_2) = w_H(\widetilde{y}) + w_H(\widetilde{c})$$

and

$$m_2 := d_H(\widetilde{y}, a_P(\widetilde{y}) - c_2) = d_H(\widetilde{y}, \widetilde{c} - c_2) = w_H(\widetilde{y}),$$

and hence $n_2 < m_2$. By Proposition 1

$$G_{(a_H, c_i)}(s) = s^{n_i} + \sum_{y \in \mathbb{F}_q^n, y \neq \widetilde{y}} s^{d_H(y, a_H(y) - c_i)}$$

and

$$G_{(a_P, c_i)}(s) = s^{m_i} + \sum_{y \in \mathbb{F}_q^n, y \neq \widetilde{y}} s^{d_H(y, a_P(y) - c_i)}.$$

As $T_{(a_H,a_P,c_i)}(s) = G_{(a_H,c_i)}(s) - G_{(a_P,c_i)}(s)$, $i = 1, 2$, and $a_H(y) = a_P(y)$ for all $y \neq \widetilde{y}$, we obtain

$$T_{(a_H,a_P,c_1)}(s) = s^{n_1} - s^{m_1}$$

and

$$T_{(a_H,a_P,c_2)}(s) = s^{n_2} - s^{m_2}.$$

Therefore $T_{(a_H,a_P,c_2)}(s) < 0 < T_{(a_H,a_P,c_1)}(s)$ for all $0 < s < 1$ and the result follows from Theorem 2

## 10.5 Proof of Theorem 8

Let $R(s_1), \ldots, R(s_M)$ be the decision regions of the decoder $a : \mathbb{R}^N \to S$. Consider a partition

$$\left\{ \widetilde{R}(s_i), \widetilde{R}(s_j) \right\}$$

of $R(s_i) \cup R(s_j)$, different from the partition $\{R(s_i), R(s_j)\}$, and such that $\widetilde{R}(s_i) = R(s_i) \cup S_j$ for some open subset $S_j \subseteq R(s_j)$ with $s_j \notin S_j$. It is obvious that such a partition exists since the decision regions of $a$ has non-empty interior. Under those conditions we have that $\widetilde{R}(s_j) = R(s_j) - S_j$.

Let

$$\widetilde{a} : \mathbb{R}^N \to S$$

be the decoder of $S$ determined by the decision regions

$$\left\{ R(s_1), \ldots, \widehat{R(s_i)}, \ldots, \widehat{R(s_j)}, \ldots, R(s_M) \right\} \cup \left\{ \widetilde{R}(s_i), \widetilde{R}(s_j) \right\}.$$

Since $\tau = s_j - s_i$ admits a unique solution on $\Delta S$, the same holds for $-\tau = s_i - s_j$ and we find the following:

$$G_a(s_j - s_i) = \int_{R(s_j)} p(y|\, s_i) P(s_i)\, dy,$$

$$G_{\widetilde{a}}(s_j - s_i) = \int_{R(s_j) - S_j} p(y|\, s_i) P(s_i)\, dy,$$

$$G_a(s_i - s_j) = \int_{R(s_i)} p(y|\, s_j) P(s_j)\, dy,$$

and

$$G_{\widetilde{a}}(s_i - s_j) = \int_{R(s_i) \cup S_j} p(y|\, s_j) P(s_j)\, dy.$$

It follows that

$$T_{(a,\widetilde{a})}\left(s_j - s_i\right) = \int_{S_j} p\left(y|\, s_i\right) P\left(s_i\right) dy > 0$$

and

$$T_{(a,\widetilde{a})}\left(s_i - s_j\right) = -\int_{S_j} p\left(y|\, s_j\right) P\left(s_i\right) dy < 0,$$

as desired.

# 11 Appendix 2: Details about "Hello World" Enconding Scheme

In the introduction of this work we simulated the transmition of the scale-of-grey image of the words "Hello World" (Figure 2) through a binary memoryless channel with crossover probability $p = 0.3$ and decoded the received word twice, once using the ML decoder and once using a $P$-NN decoder (Figure 4). The poset $P$ we used was the total order defined by the relations $1 \preceq 2 \preceq \ldots \preceq 7$. We now describe in details the codification process.

The code itself is a $[7;4]_2$ binary Hamming code, but in the encoding process not only the code as a subset is important, but also the particular color that is attributed to each codeword. The choice of the encoding is intimately related to the nature of the information and the characteristics of the $P$-NN decoder. We assumed that in the transmitted images the darker tones of gray carries the more important information, the tones used in the letters. Since the $P$-decoder is more susceptible to errors in the last coordinates we associated the darker tons of gray to codewords that has nonzero entries in the last coordinates (7 and 6), the middle range of grays to the codewords that has non-zero coordinates in the intermediate coordinates ($5, 4$ and 3) and the lighter tones of grays in the remaining positions. The actual association is shown in the following table, where the tones of gray are described by the scale in the RGB palette. Since the actual meaning of each information is relevant to decide where to place it as a codeword, we may say we are adopting a message-wise UEP encoding scheme.

| RGB | codeword $c_i$ | value $\mu(c_i)$ |
|---|---|---|
| $(101, 101, 101)$ | $c_{15} = 1111111$ | 1.00 |
| $(102, 102, 102)$ | $c_{14} = 0001111$ | 0.90 |
| $(103, 103, 103)$ | $c_{13} = 0010011$ | 0.89 |
| $(104, 104, 104)$ | $c_{12} = 1100011$ | 0.88 |
| $(105, 105, 105)$ | $c_{11} = 1010101$ | 0.87 |
| $(106, 106, 106)$ | $c_{10} = 0100101$ | 0.86 |
| $(107, 107, 107)$ | $c_9 = 0111001$ | 0.85 |
| $(108, 108, 108)$ | $c_8 = 1001001$ | 0.84 |
| $(109, 109, 109)$ | $c_7 = 0110110$ | 0.83 |
| $(110, 110, 110)$ | $c_6 = 1000110$ | 0.82 |
| $(187, 187, 187)$ | $c_5 = 1011010$ | 0.50 |
| $(188, 188, 188)$ | $c_4 = 0101010$ | 0.40 |
| $(189, 189, 189)$ | $c_3 = 0011100$ | 0.30 |
| $(190, 190, 190)$ | $c_2 = 1101100$ | 0.20 |
| $(191, 191, 191)$ | $c_1 = 1110000$ | 0.10 |
| $(192, 192, 192)$ | $c_0 = 0000000$ | 0.00 |

Considering this encoding and the values listed in the previous table, we can list all the 32 polynomials $G_{a_P}(c_i)$ and $G_{a_H}(c_i)$, $i = 0, 1, \ldots, 15$, associated to the expected loss functions $\mathbb{E}(a_P, \mu)$ e $\mathbb{E}(a_H, \mu)$ respectively. We remark (without proving) that for the Hamming case the polynomial $G_{a_H}(\tau)$ depends only on the weight $w_H(\tau)$ hence it is sufficient to know the polynomials $G_{a_H}(c_0)$, $G_{a_H}(c_{13})$, $G_{a_H}(c_{14})$ and $G_{a_H}(c_{15})$:

$$\begin{aligned} G_{a_H}(c_0) &= 16 + 112s \\ G_{a_H}(c_{13}) &= 48s^2 + 16s^3 + 64s^4 \\ G_{a_H}(c_{14}) &= 64s^3 + 16s^4 + 48s^5 \\ G_{a_H}(c_{15}) &= 112s^6 + 16s^7 \end{aligned}$$

As for $\mathbb{E}(a_P, \mu)$ we have that:

$$\begin{aligned} G_{a_P}(c_0) &= 16 + 39s + 39s^2 + 25s^3 + 9s^4 \\ G_{a_P}(c_1) &= 25s + 57s^2 + 39s^3 + 7s^4 \\ G_{a_P}(c_2) &= 10s + 42s^2 + 54s^3 + 22s^4 \end{aligned}$$

$$
\begin{aligned}
G_{a_P}(c_3) &= 6s + 22s^2 + 42s^3 + 42s^4 + 16s^5 \\
G_{a_P}(c_4) &= 7s + 39s^2 + 57s^3 + 25s^4 \\
G_{a_P}(c_5) &= 9s + 25s^2 + 39s^3 + 39s^4 + 16s^5 \\
G_{a_P}(c_6) &= 16s^2 + 42s^3 + 42s^4 + 22s^5 + 6s^6 \\
G_{a_P}(c_7) &= 22s^3 + 54s^4 + 42s^5 + 10s^6 \\
G_{a_P}(c_8) &= 10s + 42s^2 + 54s^3 + 22s^4 \\
G_{a_P}(c_9) &= 6s + 22s^2 + 42s^3 + 42s^4 + 16s^5 \\
G_{a_P}(c_{10}) &= 16s^2 + 39s^3 + 39s^4 + 25s^5 + 9s^6 \\
G_{a_P}(c_{11}) &= 25s^3 + 57s^4 + 39s^5 + 7s^6 \\
G_{a_P}(c_{12}) &= 16s^2 + 42s^3 + 42s^4 + 22s^5 + 6s^6 \\
G_{a_P}(c_{13}) &= 22s^3 + 54s^4 + 42s^5 + 10s^6 \\
G_{a_P}(c_{14}) &= 7s^3 + 39s^4 + 57s^5 + 25s^6 \\
G_{a_P}(c_{15}) &= 9s^3 + 25s^4 + 39s^5 + 39s^6 + 16s^7
\end{aligned}
$$

In the following figure we can see the graphs of the differences $T_{(a_P,a_H,c)}(s)$.

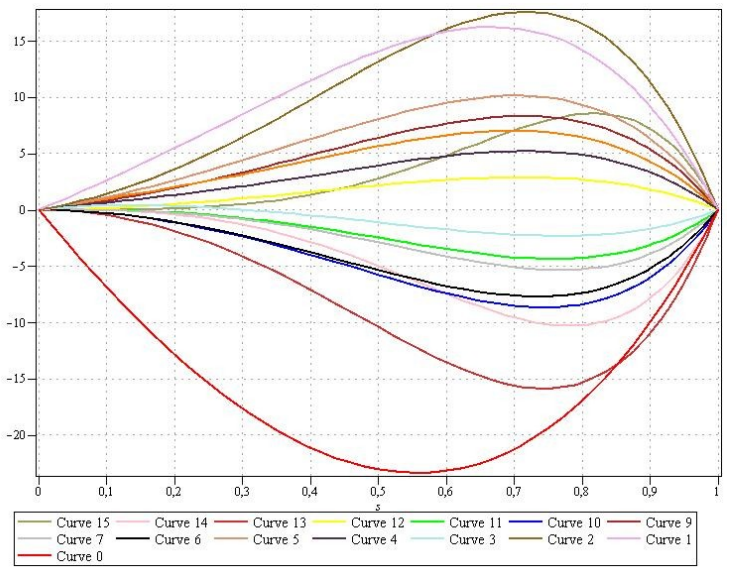

Figure 6: Difference between overall expected functions.

Computing the difference between the overall expected functions relative to the value function $\mu$ as a function of $s$ we get:

$$\mathbb{E}_{(a_P,a_H)}(\mu) = 27.10s - 58.34s^2 - 58.79s^3 + 58.97s^4 + 40.33s^5 - 9.27s^6.$$

The graph bellow shows us that for $s > 0.4$ (equivalently, for $p > 0.29$) decoder $a_P$ performs better then $a_H$.

Now we come back to the "Hello World" picture. Since the darker tones of gray are represented by codewords that have at least one of the last two coordinates with non-zero entry and since the probability of occurring an error in one of the first five entries is higher than in the last two ones, when you transmit a dark gray message, there is a higher probability that the $P$-NN decoder decodes the received message as a dark tone of gray that may be not the correct one, but looks like the original message (see Figure 4). In other words, the last two coordinates are more protected than the others for decoding in scale-of-gray. In this sense we are making a bit-wise UEP decoding.

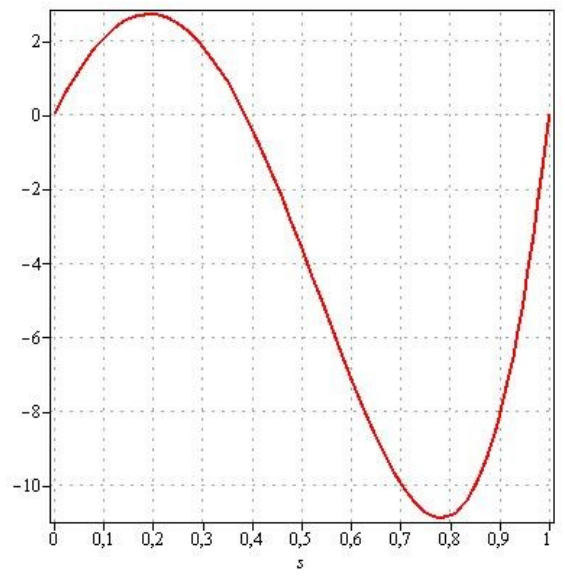

Figure 7: Graph of $\mathbb{E}_{(a_P,a_H)}(\mu)$ as a function of $s$.

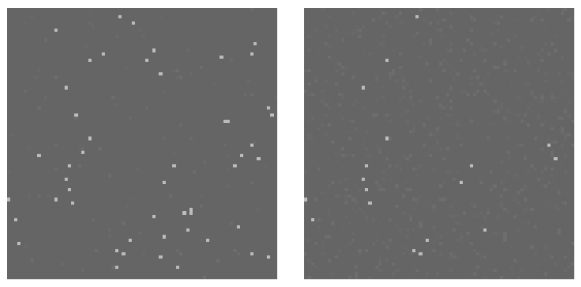

Figure 8: Each image contains 6400 pixels. The original message was the dark gray $(101, 101, 101)$ in RGB; on the left we used ML decoding and on the right the $P$-NN decoding.

Of course a repetition code could attain similar results, but in order to get a similar quality of the "Hello World" picture under severe transmission conditions (crossover probability $0.3 \le p < 0.45$), the rate of information would be much smaller than the rate achieved in this case. In Figure 9 we can see that even under a very high error probability ($p = 0.4$ and $p = 0.43$) that it is possible to grasp something of the original message.

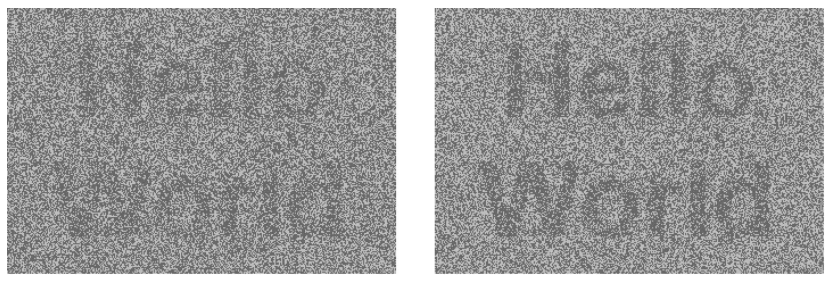

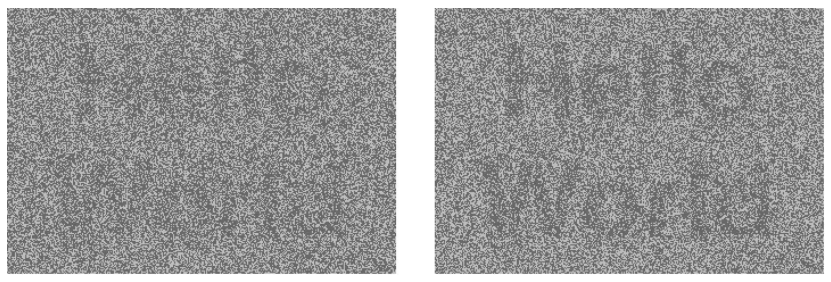

Figure 9: Image "Hello World" decoded after being transmitted through a BSMC with crossover probability $p = 0.4$ and $p = 0.43$ respectively.

**Acknowledgment** Significant part of this work was developed during the summer of 2011, where one of the authors (M. Firer) was at the Centre Interfacultaire Bernoulli at EPFL, Lausanne, Switzerland, supported by the Swiss National Science Foundation, to which he deeply thanks. The second author is currently pursuing the Ph.D. degree in the Department of Mathematics, UEM, Brazil. During 2011, L. Rifo is visiting professor at CIMFAV, at the University of Valparaiso, Chile, supported by Proyecto MEL-Conicyt.